%% file: main.tex
\documentclass[letterpaper]{article} % DO NOT CHANGE THIS
\usepackage{aaai24}  % DO NOT CHANGE THIS
\usepackage{times}  % DO NOT CHANGE THIS
\usepackage{helvet}  % DO NOT CHANGE THIS
\usepackage{courier}  % DO NOT CHANGE THIS
\usepackage[hyphens]{url}  % DO NOT CHANGE THIS
\usepackage{graphicx} % DO NOT CHANGE THIS
\usepackage{natbib}  % DO NOT CHANGE THIS AND DO NOT ADD ANY OPTIONS TO IT
\usepackage{caption} % DO NOT CHANGE THIS AND DO NOT ADD ANY OPTIONS TO IT
\usepackage{algorithm}
\usepackage{algorithmic}

\usepackage{amsmath}
\usepackage{mathtools}
\usepackage{amsthm}
\usepackage{bm}
\usepackage{threeparttable}  
\usepackage{multirow}
\usepackage{natbib}
\usepackage{subfigure}
\usepackage{bbding}
\usepackage{booktabs}
\usepackage{colortbl}
\usepackage{enumitem}
\usepackage{xcolor}
\usepackage{amssymb}
\usepackage{url}

\usepackage{newfloat}
\usepackage{listings}
\DeclareCaptionStyle{ruled}{labelfont=normalfont,labelsep=colon,strut=off} % DO NOT CHANGE THIS
\floatstyle{ruled}
\newfloat{listing}{tb}{lst}{}
\floatname{listing}{Listing}

\title{STEM: Unleashing the Power of Embeddings for Multi-task Recommendation}
\author {
    Liangcai Su \equalcontrib\textsuperscript{\rm 1},
    Junwei Pan \equalcontrib\textsuperscript{\rm 2},
    Ximei Wang\textsuperscript{\rm 2},
    Xi Xiao\textsuperscript{\rm 1}\thanks{Corresponding author.},
    Shijie Quan\textsuperscript{\rm 2},
    Xihua Chen\textsuperscript{\rm 2},
    Jie Jiang\textsuperscript{\rm 2},
}
\affiliations {
    \textsuperscript{\rm 1}Shenzhen International Graduate School, Tsinghua University \\
    \textsuperscript{\rm 2}Tencent Inc.\\
    sulc21@mails.tsinghua.edu.cn, xiaox@sz.tsinghua.edu.cn,\\
    \{jonaspan, messiwang, justinquan, tinychen, zeus\}@tencent.com
}

\usepackage{bibentry}
\begin{document}

\maketitle

\begin{abstract}
Multi-task learning (MTL) has gained significant popularity in recommender systems as it enables simultaneous optimization of multiple objectives. 
A key challenge in MTL is negative transfer, but existing studies explored negative transfer on all samples, overlooking the inherent complexities within them. 
We split the samples according to the relative amount of positive feedback among tasks.
Surprisingly, negative transfer still occurs in existing MTL methods on samples that receive comparable feedback across tasks. 
Existing work commonly employs a shared-embedding paradigm, limiting the ability of modeling diverse user preferences on different tasks. 
In this paper, we introduce a novel Shared and Task-specific EMbeddings (STEM) paradigm that aims to incorporate both shared and task-specific embeddings to effectively capture task-specific user preferences.
Under this paradigm, we propose a simple model STEM-Net, which is equipped with an All Forward Task-specific Backward gating network to facilitate the learning of task-specific embeddings and direct knowledge transfer across tasks.
Remarkably, STEM-Net demonstrates exceptional performance on comparable samples, achieving positive transfer.
Comprehensive evaluation on three public MTL recommendation datasets demonstrates that STEM-Net outperforms state-of-the-art models by a substantial margin. Our code is released at 
\textcolor{magenta}{\url{https://github.com/LiangcaiSu/STEM}}.

% Multi-task learning (MTL) has gained significant popularity in recommendation systems as it enables the simultaneous optimization of multiple objectives. 
% A key challenge in MTL is negative transfer, where the performance of certain tasks deteriorates due to conflicts between tasks. 
% Existing research has explored negative transfer by treating all samples as a whole, overlooking the inherent complexities within them. 
% To this end, we delve into the intricacies of samples by splitting them based on the relative amount of positive feedback among tasks.
% Surprisingly, negative transfer still occurs in existing MTL methods on samples that receive comparable feedback across tasks. 
% It is worth noting that existing methods commonly employ a shared-embedding paradigm, and we hypothesize that their failure can be attributed to the limited capacity of modeling diverse user preferences across tasks using such universal embeddings. 
\end{abstract}

\input{sections/1_introduction}

\input{sections/2_related_work}

\input{sections/3_analysis}
\input{sections/4_model}

\input{sections/5_experiment}
\input{sections/6_conclusion}

\newpage
\section{Acknowledgments}
This work was supported in part by the National Natural Science Foundation of China (61972219), the Overseas Research Cooperation Fund of Tsinghua Shenzhen International Graduate School (HW2021013). The authors would like to thank Zhutian Lin for his help with the contradictory user preference analysis.
\bibliography{aaai24}
\newpage

\end{document}

%% file: sections/1_introduction.tex
\section{Introduction}

Recently, multi-task learning has drawn great interest in recommender systems since they are able to optimize multiple objectives (\textit{e.g.}, \texttt{Like}, \texttt{Share}, \texttt{Finish}) simultaneously. 
The effectiveness of multi-task recommendation (MTR) depends on leveraging knowledge from other tasks to \textit{help} the learning of each task. 
However, prior works have identified negative transfer~\cite{negative_transfer_2010} and seesaw phenomenon~\cite{ple}, whereby multi-task learning models may not always outperform single-task models. 
To address negative transfer,
MMoE~\cite{mmoe} and PLE~\cite{ple} incorporate \textit{task-specific} gate networks and experts, respectively. 
Additionally, techniques such as gradient clipping~\cite{gradientsurgery,chen2018gradnorm, MetaBalanceWWW2022} and task-aware optimizer~\cite{gradientsurgery, chen2018gradnorm, adatask2023} have also been employed to alleviate gradient conflicts.

\input{fig/intro_buckets}

\input{fig/related_work}

Existing methods tend to treat all samples in a task as a whole, overlooking the inherent intricacies within them. 
Consequently, it remains unclear \textit{where negative transfer occurs}. 
To address this concern, we split the test set of TikTok, a public MTL recommendation dataset, into three subsets according to the relative amount of feedback between task \texttt{Finish} and \texttt{Like}: \texttt{Finish}-Overwhelming, Comparable, and \texttt{Like}-Overwhelming. 
We then evaluate the performance of two popular MTL models, MMoE and PLE, and the single-task model on task \texttt{Like} over these subsets, since it has much fewer positive samples and therefore suffers from negative transfer.
% It is noteworthy that the single-task \texttt{Like} model comprises significantly fewer positive samples, making it susceptible to negative transfer in MTL.
As shown in Figure~\ref{fig:intro_buckets}, compared to the single task model, both MMoE and PLE demonstrate a notable performance lift on the \texttt{Finish}-Overwhelming subset by incorporating additional knowledge of task \texttt{Finish} to resolve the feedback sparsity on \texttt{Like}. 
They also have a slight performance boost on the \texttt{Like}-Overwhelming subset where task \texttt{Like} itself can already learn well, but stills benefits from the complementary knowledge from \texttt{Finish}.
However, to our surprise, on the subset that receives comparable positive feedback from both tasks, the performance of existing methods is \textit{inferior} to that of the single-task model.

The performance drop of existing MTL methods on the comparable subset prompts us to reconsider the design of MTL recommendation models. 
Intuitively, users may possess diverse and sometimes even conflicting preferences over items across various tasks, which become more pronounced when there are sufficient signals from multiple tasks, as on the comparable subset here. 
The preference of users is captured through user and item embeddings in recommendation models.
However, existing MTL methods, including MMoE~\cite{mmoe} and PLE~\cite{ple}, all follow a \emph{shared-embedding paradigm}.
That is, they learn a universal embedding for each user and item, shared across tasks.
Such a paradigm is only able to capture a single preference of users, hindering the ability to capture the preference divergence across tasks.

Motivated by the limitations of the shared-embedding paradigm, this paper introduces a \underline{S}hared and \underline{T}ask-specific \underline{EM}beddings (STEM) paradigm. 
STEM aims to \emph{incorporate both shared and task-specific embeddings} to learn common and task-specific user preference.
Under this paradigm, we design a simple model, namely STEM-Net, which begins by introducing the shared and task-specific embedding tables.
STEM-Net then constructs a set of experts, where each expert is either shared across tasks, utilizing only shared embeddings, or task-specific, utilizing embeddings specific to a particular task.
Furthermore, STEM-Net employs an All Forward Task-specific Backward gating network for each task tower, which a) \emph{receive forward from shared and all task-specific experts} and b) employs a \emph{stop-gradient operation on the experts of the other tasks} to learn task-specific embeddings by preventing a task's gradients from updating embeddings of the other tasks. 

We conduct extensive evaluation of STEM-Net on three subsets of the TikTok dataset. 
STEM-Net consistently outperforms both MMoE and PLE on the \texttt{Finish} and \texttt{Like}-Overwhelming subsets.
Moreover, STEM-Net exhibits remarkable performance in the comparable subset, surpassing the Single-Task \texttt{Like} model. 
% This achievement \emph{signifies the successful mitigation of negative transfer}, marking a notable advancement in the field of MTL.
Furthermore, we assess STEM-Net on three public MTL recommendation datasets. 
In all cases, STEM-Net surpasses state-of-the-art models by a substantial margin, further validating its effectiveness. 
We outline our contributions as follows:

\begin{itemize}
    \item We delve into the intricacies of samples to investigate the negative transfer in MTL recommendation models. 
    Surprisingly, our investigation reveals the presence of negative transfer on samples that receive comparable feedback from both tasks.
    \item We propose a novel paradigm called STEM (Shared and Task-specific EMbeddings) for multi-task recommendation. 
    Under this paradigm, we design a simple model called STEM-Net equipped with an All Forward Task-specific Backward gating network to facilitate direct knowledge transfer across tasks while learning task-specific embeddings.
    \item We conduct comprehensive experiments and ablation studies on three MTL recommendation datasets and provide compelling evidence of STEM-Net's effectiveness.
    In online A/B testing, STEM-Net achieves significant improvement in GMV (Gross Mechanise Value) over the base model (MMoE), and it has been successfully deployed in Tencent's online advertising platform.
\end{itemize}

%% file: fig/intro_buckets.tex
%%%%%%%%%%%%%%%%%%%%%%%%%%%%%%%%%%%%%%%%%%%%%%%%%%%%%%%%%%
\begin{figure}[tp!]
	\centering
	\includegraphics[width=1.0\linewidth]{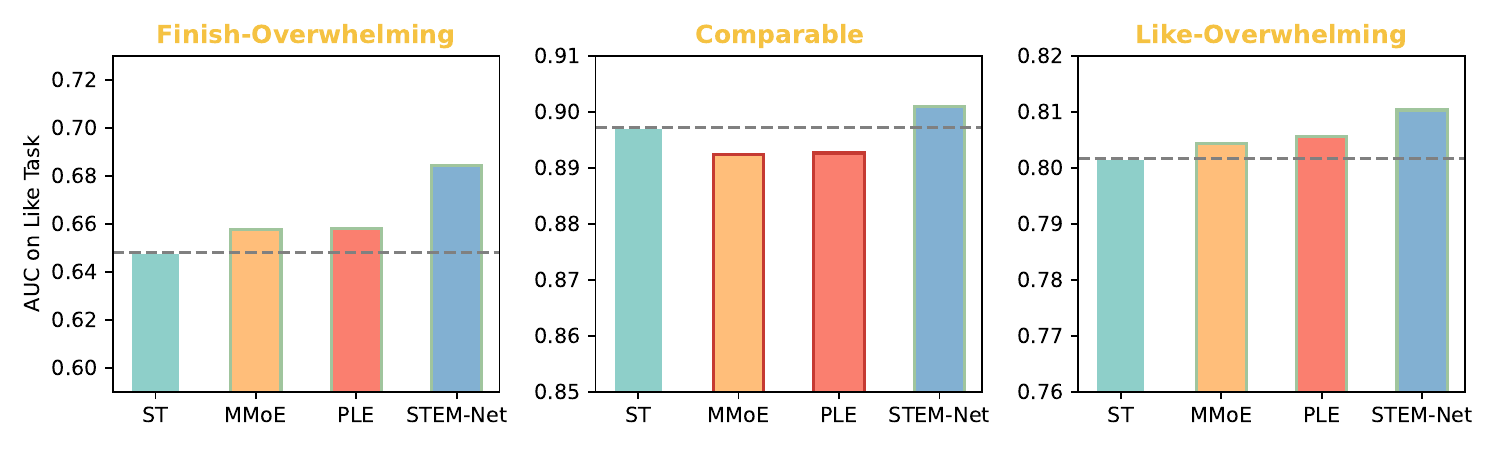}
	% \caption{AUC gains of MTL Models under different sub-buckets compared to a single-task model. \textcolor[HTML]{DE8344}{+} means the number of positive samples."}
        \caption{Existing MTL models such as MMoE and PLE suffer from negative transfer on the comparable subset of TikTok test samples, while STEM-Net achieves positive transfer. 
        STEM-Net also outperforms MMoE and PLE on task-overwhelming subsets.
        ST: Single-Task.}
	\label{fig:intro_buckets}
\end{figure}
%%%%%%%%%%%%%%%%%%%%%%%%%%%%%%%%%%%%%%%%%%%%%%%%%%%%%%%%%%

%% file: fig/related_work.tex
%%%%%%%%%%%%%%%%%%%%%%%%%%%%%%%%%%%%%%%%%%%%%%%%%%%%%%%%%%
\begin{figure*}[t!]
	\centering
\includegraphics[width=\linewidth]{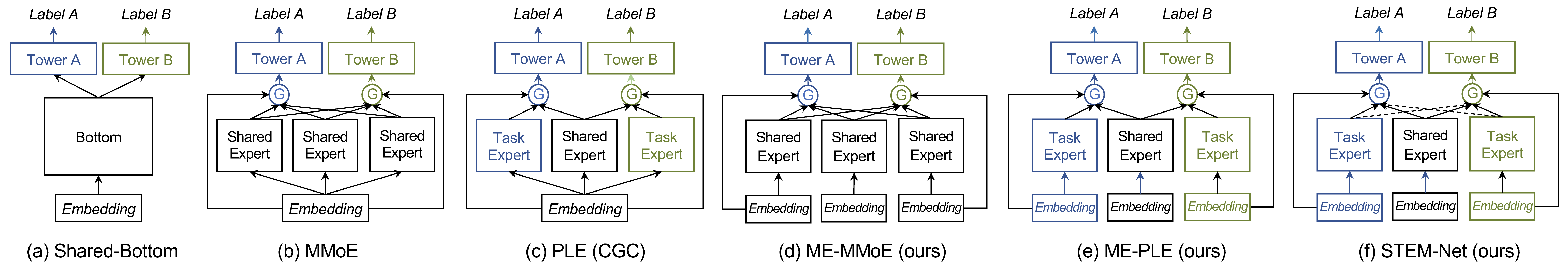}
	\caption{Comparison between representative MTL models and our proposed STEM-Net. Dot lines denote connections with stop-gradient operation.}
	\label{fig:related_work}
\end{figure*}
%%%%%%%%%%%%%%%%%%%%%%%%%%%%%%%%%%%%%%%%%%%%%%%%%%%%%%%%%%

%% file: sections/2_related_work.tex
\section{Related Work}
\label{sec:related_work}
\subsection{Evolution of the MTL Model Architectures}

In this section, we provide a comprehensive review of multi-task recommendation models that are based on the widely adopted \textit{Embedding-Expert-Gate-Tower} architecture~\cite{shared_bottom,mmoe,ple,crossdistill2022,hazimeh2021dselect,mose_kdd2020}.
This architecture comprises four key modules: \textit{Embeddings}, which employ dense vectors to represent sparse features; \textit{Experts}, which extract knowledge from the input features; \textit{Gates}, which aggregate the outputs of the experts using attentive weights; and \textit{Towers}, which are task-specific classifiers.
Figure~\ref{fig:related_work} illustrates representative models at various levels of task specificity.
Notably, recent works have witnessed a gradual shift in the placement of task-specific modules, from top modules such as towers~\cite{shared_bottom, Cross-Stitch} or gates~\cite{mmoe, mtfwfm2019} toward bottom modules such as experts~\cite{ple}.

\textbf{Tower-level task-specific models}, where each task has an independent tower and is updated only by its own loss, as shown in Figure~\ref{fig:related_work}(a).
All remaining parameters are shared across tasks.
Shared-Bottom~\cite{shared_bottom} is a representative of these models, also known as hard parameter sharing. 
OMoE~\cite{mmoe}, Cross-stitch~\cite{Cross-Stitch} are variants of Shared-Bottom with shared experts (\textit{i.e}. bottom). 

\textbf{Gate-level task-specific models}, which employ a gating network to each task to weight the outputs of experts.
MMoE is a representive of these models, as shown in Figure~\ref{fig:related_work}(b). 
Dselect-k~\cite{hazimeh2021dselect}, MoSE~\cite{mose_kdd2020} and MT-FwFM~\cite{mtfwfm2019} also fit into this category. 
All parameters in the modules below the gates, including experts and embeddings, are shared across tasks.

\textbf{Expert-level task-specific models}, where each task has its own experts on the basis of the Tower/Gate-level task-specific model. 
PLE~\cite{ple} is a representative model, which utilizes both task-specific and shared experts. 
Numerous subsequent works follow this design including PFE~\cite{pfe_www2022}, MLPR~\cite{MLPR_WWW2022} and TAML~\cite{liu2023taml}.  
Moreover, some methods based on sparse routings, such as SNR~\cite{snr} and CSRec~\cite{CSRecWWW2022}, are also Expert-level task-specific models since some parameters of their experts are task-specific.
In these models, even though parameters of some experts are task-specific, the embeddings are still shared across tasks.

In contrast to the above studies, we focus on \textbf{Embedding-level task-specific models}. We will discuss this in detail in the method section.

%% file: sections/3_analysis.tex
\input{fig/overview}

\section{Delve into Negative Transfer in MTL Recommenders}

In MTL recommendation, negative transfer refers to the phenomenon where knowledge or information learned from one task adversely affects the performance of another task.
%Understanding and mitigating negative transfer is crucial for improving the effectiveness and efficiency of recommender systems.
Existing research primarily focuses on investigating negative transfer across all samples as a whole, often overlooking the inherent intricacies within samples. 
To illustrate this point, let us consider an MTL setting involving two tasks, denoted as task \texttt{A} and \texttt{B}. 
Samples with limited positive feedback from \texttt{A} may intuitively benefit from MTL by receiving additional feedback from \texttt{B}. 
Besides, samples with abundant positive feedback from \texttt{A} can also get performance lift by incorporating complementary feedback from \texttt{B}. 
However, when a comparable amount of feedback exists for both tasks \texttt{A} and \texttt{B}, negative transfer may occur due to the possible contradictory user preferences over items between two tasks.

To verify this hypothesis, we can split the testing samples into three subsets according to the relative amount of (expected) positive feedback between task \texttt{A} and \texttt{B}: \texttt{A}-Overwhelming subset $\mathcal{D}_\text{A-O}$ and \texttt{B}-Overwhelming subset $\mathcal{D}_\text{B-O}$, each consisting of samples with \emph{overwhelming positive feedback} from task \texttt{A} or \texttt{B}, respectively, and Comparable subset $\mathcal{D}_\text{Comp}$ which consists of samples with \emph{comparable positive feedback} from both tasks.
In particular, we measure the amount of positive feedback for a given sample regarding each task by the \emph{expected} positive feedback from each \emph{single task model}, and then split the samples into subsets according to the gap between task-wise expected positive feedback.

We conducted an empirical analysis on the TikTok dataset, which consists of two tasks: \texttt{Like} and \texttt{Finish}. 
We aim to investigate negative transfer on the \texttt{Like} task, as it exhibits a significantly lower number of positive samples, rendering it susceptible to be dominated.
Following the above mentioned procedure, we first discretize the expected feedback of each task into 10 buckets with equal sample frequency.
The relative feedback can then be quantified by the difference in bucket indices: $b(f_A(x)) - b(f_B(x))$.
We define \texttt{Finish}-Overwhelming subset as the set of samples with a bucket index difference in the range of $[-9, -4]$, the Comparable subset with a range of $(-4, 6]$ and \texttt{Like}-Overwhelming with a range of $(6, 9]$.
We evaluate the performance of two existing MTL models, MMoE and PLE, and the \texttt{Like} single task model, on the \texttt{Like} task.
As depicted in Figure~\ref{fig:intro_buckets}, both MMoE and PLE demonstrate improved performance when there is overwhelming feedback from either task.
However, both models exhibit inferior performance compared to the single task model on subset $\mathcal{D}_\text{Comp}$.
We attribute the performance drop of MMoE and PLE to that there may be contradictory user preference over items between \texttt{Like} and \texttt{Finish} on the comparable subset. Shared embedding methods, such as MMoE and PLE, is incapable to capture such contradictory preference.
We'll investigate contradictory user preference in the Performance Evaluation Section.
In the following, we'll present our STEM paradigm and STEM-Net under this paradigm.

%% file: fig/overview.tex
%%%%%%%%%%%%%%%%%%%%%%%%%%%%%%%%%%%%%%%%%%%%%%%%%%%%%%%%%%
\begin{figure*}[t!]
	\centering
\includegraphics[width=0.95\linewidth]{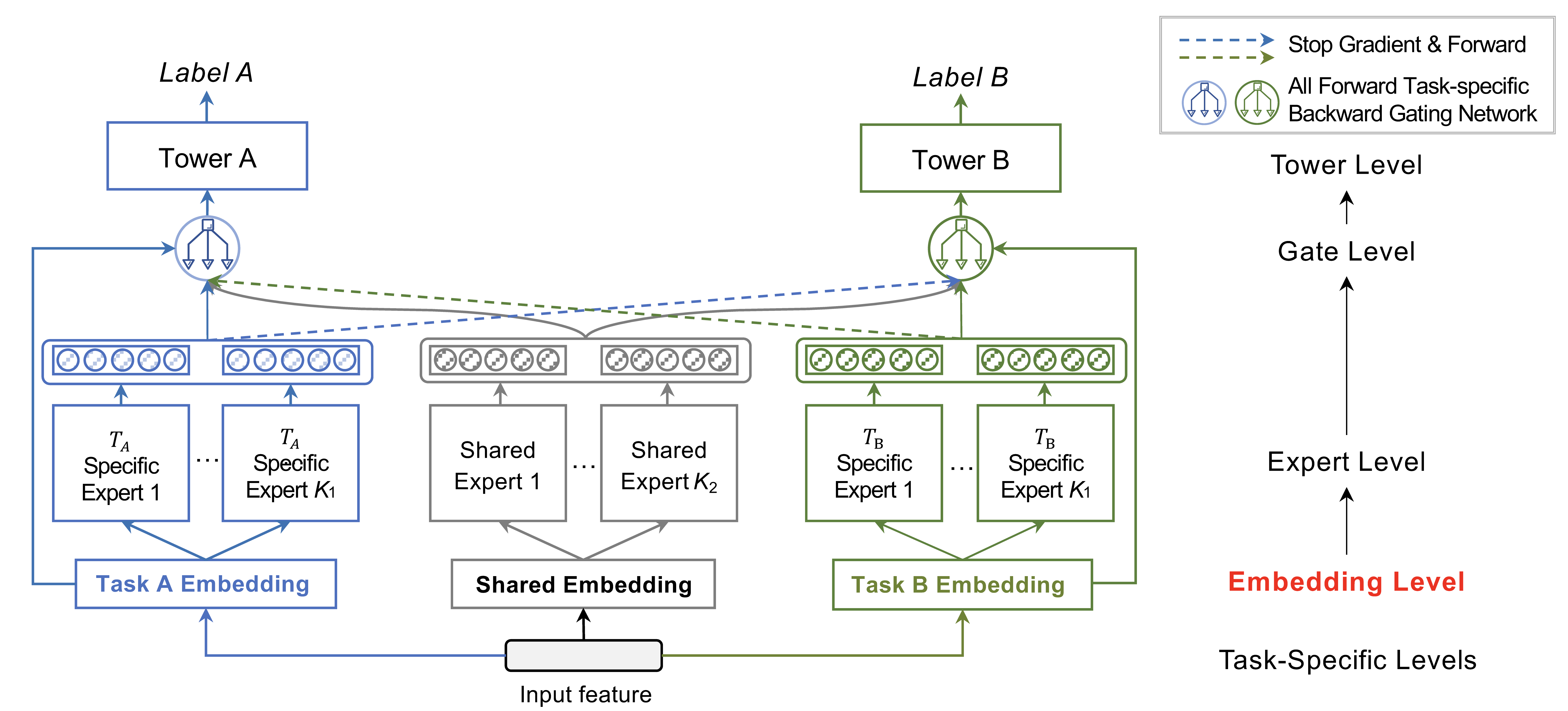}
	\caption{Overview of STEM-Net.}
	\label{fig:overview}
\end{figure*}
%%%%%%%%%%%%%%%%%%%%%%%%%%%%%%%%%%%%%%%%%%%%%%%%%%%%%%%%%%

%% file: sections/4_model.tex
\section{Method}

\subsection{The STEM Paradigm}
\label{sec:STEM_paradigm}

In recommendation, each sample consists of $M$ active features, i.e.,  $x=\{x_1, x_2,\cdots, x_M\}$, where $M$ denotes the number of fields and  $x_i$ represents the active feature of the $i$-th field. 
All existing methods follow a shared-embedding paradigm, that is, they have only one shared embedding table, and learn one shared embedding for each feature.

Under STEM paradigm, in addition to the shared embedding table $E^S \in \mathcal{R}^{N \times K}$ in the shared-embedding paradigm, we also employ $T$ task-specific embedding tables $\{E^1, \dots, E^t, \dots, E^T\}$, one for each task $t$.
Here $N$ denotes the total number of features across all fields, and $K$ denotes the embedding dimension. 
For a given feature $x_i$, we get the task-specific embeddings $\{\bm{v}_i^t\}$ as well as the shared embedding $\bm{v}_i^S$ as follows :
\begin{equation}
    \begin{aligned}
        \bm{v}^{t}_i & = \texttt{Lookup}(x_i, E^t),  \\  %= E^t[i,:]
        \bm{v}^{S}_i & = \texttt{Lookup}(x_i, E^S). \\  %= E^S[i,:] 
    \end{aligned}
\end{equation}

We concatenate the task-specific and shared embeddings across all active features as follows:
\begin{equation}
    \begin{aligned}
        \bm{h}_0^t & = [\bm{v}^{t}_1, \dots, \bm{v}^{t}_i, \dots, \bm{v}^{t}_M], \\ 
        \bm{h}_0^S & = [\bm{v}^{S}_1, \dots, \bm{v}^{S}_i, \dots, \bm{v}^{S}_M], \\ 
    \end{aligned}
\end{equation}
where the concatenated task-specific embedding $\bm{h}_{0}^{t}$ and shared-embedding $\bm{h}_0^S$ are the input to the experts, as discussed below.

\subsection{The STEM-Net Model}

Under STEM paradigm, our proposed STEM-Net further employ Shared \& Task-Specific Experts and an All Forward Task-specific Backward gating network to facilitate the learning of task-specific embeddings and knowledge transfer across tasks.

\subsubsection{Shared \& Task-Specific Experts}

% Multi-task models have been utilizing experts to capture shared or task-specific to model the common and different knowledge of multiple tasks.
Following PLE~\cite{ple}, we employ shared and task-specific experts.
However, each shared or task-specific expert group is equipped with an independent embedding table, so as to prevent any parameters interference.
In particular, the shared experts only takes $\bm{h}_0^S$ as the input, which consists of embeddings from the $E^S$. 
Experts for task $t$ only takes $\bm{h}_0^t$ as the input, consisting of embeddings from $E^t$.
For brevity, we assume that each task-specific expert group and shared expert group contains $K_1$ and $K_2$ experts respectively. Each expert is a multi-layer perceptrons (MLPs) over the input: 
\begin{equation}
\begin{aligned}
        \bm{h}^{t}_{i} &=  \texttt{MLP}^{t}_{i}(\bm{h}_0^t), \forall i=1,\dots,K_1, \\ 
        \bm{h}^{S}_{j} & = \texttt{MLP}^{S}_{j}(\bm{h}_0^S), \forall j=1,\dots,K_2, 
\end{aligned}
\end{equation}
where $\bm{h}^{t}_{i}$ represents output of the $i$-th expert for task $t$ and $\bm{h}^{S}_{j}$ represents the output of the $j$-th shared expert. 

\subsubsection{All Forward Task-specific Backward Gating Network}
The gating mechanism aims to integrate outputs from experts for each task's tower. 
In STEM-Net, we'd like to receive the outputs from all experts, including both shared and task-specific ones, to maximize knowledge transfer, while at the same time learn task-specific embeddings, so as to capture task-specific preference.
We achieve this by designing an All Forward Task-specific Backward gating network, that connects the tower of each task to all experts, with a stop gradient operation on the experts of the other tasks.
Formally, the output of  the gating network for task $t$ is formalized as:
\begin{equation}
    \begin{aligned}
        \bm{o}^{t} & =  \sum_{i}^{K_1} \bm{g}^{t\rightarrow{t}}_{i} \bm{h}^{t}_{i} +  \sum_{i}^{K_2} \bm{g}^{S\rightarrow{t}}_{i} \bm{h}^{S}_{i}  \\ 
                & + \sum_{t' \in {\mathcal{T}}, t'\neq{t}}\sum_{i}^{K_1} \bm{g}^{t'\rightarrow{t}}_{i}\texttt{SG}(\bm{h}^{t'}_i),
    \end{aligned}
\end{equation}
where $\texttt{SG}(\cdot)$ is the stop gradient operator, and $\bm{g}^{t\rightarrow{t}}\in{\mathcal{R}^{K_1}}$ denotes the attentive weight on connections between task $t$'s corresponding tower and experts, 
$\bm{g}^{t'\rightarrow{t}}\in{\mathcal{R}^{K_1}}$ denotes the weight between the expert of task $t'$ and tower of task $t$, $\bm{g}^{S\rightarrow{t}}\in{\mathcal{R}^{K_2}}$ denotes the weight between shared experts and tower of task $t$, respectively. 
These weights are computed with a softmax over the concatenated embeddings:
\begin{equation}
    \begin{aligned}
             [\bm{g}^{t\rightarrow{t}}, \{\bm{g}^{t'\rightarrow{t}}\}, \bm{g}^{S\rightarrow{t}}] 
            & = \texttt{Softmax}(W_{g}^{t}(\bm{h}^{t}_{0}+\bm{h}^{S}_{0})), \\  
            % t'\in{T} \& t'\neq{t}
    \end{aligned}
    \label{eq:gate}
\end{equation}
where $W^{t}_{g}\in{\mathcal{R}^{d\times(K_1\times{T}+K_{2})}}$.

\input{tables/dataset_stat}

\input{tables/tiktok_qkvideo_performance}
\input{tables/kuairand1k_performance}

\subsubsection{Towers and Loss Function}
Finally, we assign each task an independent tower and obtain the final output as:
\begin{equation}
    \hat{y}_t = \sigma(\texttt{MLP}^{t}(\bm{o}^{t})),
\end{equation}
where $\sigma$ is the sigmoid function.
We choose binary cross-entropy loss function as the objective function for each task, and the final loss function is formulated as follows:
\begin{equation}
    \begin{aligned}
            \mathcal{L} % & = \sum_{i}^{|\mathcal{T}|} w_{i} \mathcal{L}_{BCE}(y_i,\hat{y}_{i}) \\
                        & = - \sum_{t}^{{T}} y_{t}\log(\hat{y}_{t}) + (1-y_{t})\log(1-\hat{y}_{t}),
    \end{aligned}
\end{equation}
where $y_{t}$ is the ground truth of task $t$.
\input{fig/adapter_comparsion}

\subsubsection{Comparison of Gating Networks}

MMoE adopts an All Forward All Backward gating network, that each task's tower receives output from all experts, and propagate the gradients to all experts, too.
All Backward makes MMoE unable to learn task-specific embeddings, even if we assign independent embedding tables for each expert, denoted as Multi-Embedding MMoE (ME-MMoE for short).

PLE adopts a Task-specific Forward Task-specific Backward gating network, where each task's tower receives outputs from shared experts and its own experts, but not from experts of the other tasks. 
Similarly, each task's tower only backward propagate gradients to the shared experts and its own experts, but not to the other tasks' experts.
Task-specific Forward makes PLE can't fully transfer the knowledge from other tasks \emph{directly} besides the shared expert, even if we assign independent embeddings for each expert (group) in PLE, denoted as ME-PLE.
The comparison of our gating network with those of MMoE and PLE is shown in Figure~\ref{fig:adapter_comparison}.

%% file: tables/dataset_stat.tex
%%%%%%%%%%%%%%%%%%%%%
\begin{table*}[]
\centering
\caption{Statistics of processed datasets.}
\resizebox{0.88\linewidth}{!}{
\begin{tabular}{c|c|c|c|c|c|c}
% \hline 
\toprule
Dataset  & \#User & \#Items & \#Samples   & \#Fields   & \#Tasks & Positive Ratio (\%)           \\ \hline
TikTok & 560K & 1800K  & 223.4M/24.8M/27.6M  &  8  & 2        & 28.31/1.60    \\ \hline
QK-Video & 970K & 760K  & 95.9M/12.0M/12.5M  &  16   & 2        & 24.01/2.03    \\ \hline
KuaiRand1K & 1K & 189K & 10.9M/0.39M/0.42M   &  32  & 8       & \begin{tabular}[c]{@{}l@{}}37.76/1.54/0.10/0.26/0.08/0.10/26.17/1.78\end{tabular} \\

\bottomrule 

\end{tabular}
}
\label{tab:datasets}
\end{table*}
%%%%%%%%%%%%%

%% file: tables/tiktok_qkvideo_performance.tex
\begin{table*}[t!]
\centering
\begin{minipage}{\textwidth}
\begin{minipage}[t]{0.49\textwidth}
\makeatletter\def\@captype{table}
\caption{Overall performance on TikTok.}
\resizebox{0.98\linewidth}{!}{
\begin{tabular}{l|cc|cc|c}
\toprule
\multirow{2}{*}{Model} & \multicolumn{2}{c|}{\texttt{Finish}} & \multicolumn{2}{c|}{\texttt{Like}} & \multirow{2}{*}{\begin{tabular}[c]{@{}c@{}}Average\\ AUC $\uparrow$\end{tabular}} \\ \cline{2-5}
                       & Logloss$\downarrow$     & AUC $\uparrow$        & Logloss$\downarrow$      & AUC$\uparrow$         &                                                                        \\ \hline
Single-Task & 0.5111 & 0.7505 & 0.0558 & 0.9058 & 0.8281 \\ \hline
Shared-Bottom & 0.5112 & 0.7504 & 0.0560 & 0.9022 & 0.8263 \\
OMoE & \textbf{0.5103} & \textbf{0.7516} & 0.0559 & 0.9029 & 0.8273 \\
MMoE & 0.5105 & 0.7511 & 0.0560 & 0.9018 & 0.8265 \\
PLE & 0.5105 & 0.7511 & 0.0560 & 0.9016 & 0.8264 \\ 
ESMM & 0.5111 & 0.7503 & 0.0564 & 0.9012 & 0.8258 \\ 
AITM & 0.5109 & 0.7506  & 0.0560  & 0.9026  & 0.8266  \\ 

\hline
ME-MMoE & 0.5114 & 0.7502 & 0.0557 & 0.9045 & 0.8274 \\
ME-PLE & 0.5120 & 0.7492 & 0.0560 & 0.9058 & 0.8275

\\
\hline

STEM-Net & {0.5104}  & 0.7513    & \textbf{0.0553}   & \textbf{0.9095} & \textbf{0.8304}    \\

\bottomrule
\end{tabular}
}
\label{tab:tiktok_performance}

\end{minipage}
\begin{minipage}[t]{0.49\textwidth}
\makeatletter\def\@captype{table}
\caption{Overall performance on QK-Video.}
\resizebox{0.98\linewidth}{!}{
\begin{tabular}{l|cc|cc|c}
\toprule

\multirow{2}{*}{Model} & \multicolumn{2}{c|}{\texttt{Click}} & \multicolumn{2}{c|}{\texttt{Like}} & \multirow{2}{*}{\begin{tabular}[c]{@{}c@{}}Average\\ AUC $\uparrow$\end{tabular}} \\ \cline{2-5}
                       & Logloss$\downarrow$     & AUC $\uparrow$        & Logloss$\downarrow$      & AUC$\uparrow$         &                                                                        \\ \hline
Single-Task & 0.2826  & 0.9234  & 0.0378  & 0.9400  & 0.9317  \\ \hline
Shared-Bottom & 0.2857  & 0.9235  & 0.0380  & 0.9389  & 0.9312  \\
OMoE & 0.2826  & 0.9238  & \textbf{0.0373}  & 0.9394  & 0.9316  \\
MMoE & 0.2813  & 0.9238  & 0.0379  & 0.9401  & 0.9319  \\
PLE & 0.2832  & 0.9238  & 0.0375  & 0.9399  & 0.9318  \\ 
ESMM & 0.2847  & 0.9208  & 0.0378  & 0.9368  & 0.9288  \\ 
AITM & 0.2836  & 0.9237  & 0.0386  & 0.9398  & 0.9318  \\ 

\hline
ME-MMoE& 0.2815  & \textbf{0.9239}  & 0.0375  & 0.9407  & 0.9323  \\
ME-PLE & 0.2818  & 0.9238  & 0.0374  & 0.9410  & 0.9324

\\
\hline
STEM-Net & 0.2816  & 0.9237  & 0.0381  & \textbf{0.9426}  & \textbf{0.9331}     \\

\bottomrule
\end{tabular}
}
\label{tab:qkvideo_performance}
\end{minipage}
\end{minipage}\

\end{table*}

%% file: tables/kuairand1k_performance.tex
\begin{table*}[t!]
\caption{Overall performance on KuaiRand1K.}
\centering
\resizebox{0.88\textwidth}{!}{
\begin{tabular}{lcccccccccc}
\toprule
Model & Task \texttt{A} & Task \texttt{B} & Task \texttt{C} & Task \texttt{D} & Task \texttt{E} & Task \texttt{F} & Task \texttt{G} & Task \texttt{H} & Avg. AUC & MTL Gain \\ \hline
Single-Task & 0.7534 & 0.9293 & 0.8294 & 0.8943 & 0.8572 & 0.8821 & 0.7650 & 0.8358 & 0.8433 & - \\ \hline
Shared-Bottom & 0.7535 & 0.9261 & 0.8162 & 0.8881 & 0.8228 & 0.7820 & 0.7642 & 0.8340 & 0.8234 & -0.0199 \\
OMoE & 0.7549 & 0.9273 & 0.8404 & 0.8923 & 0.8352 & 0.8750 & 0.7655 & 0.8349 & 0.8407 & -0.0026 \\
MMoE & 0.7541 & 0.9278 & 0.8268 & 0.8901 & 0.8591 & 0.8908 & 0.7647 & 0.8360 & 0.8437 & +0.0003 \\
PLE & 0.7537 & 0.9290 & 0.8362 & 0.8885 & 0.8449 & 0.8940 & 0.7643 & 0.8374 & 0.8435 & +0.0002 \\ \hline
ME-MMoE & \textbf{0.7555} & 0.9288 & 0.8310 & 0.8912 & 0.8500 & 0.8668 & \textbf{0.7658} & \textbf{0.8385} & 0.8410 & -0.0024 \\
ME-PLE & 0.7536 & \textbf{0.9294} & 0.8353 & \textbf{0.8970} & 0.8521 & 0.8871 & 0.7637 & 0.8381 & 0.8445 & +0.0012 \\ \hline
STEM-Net & 0.7523 & 0.9282 & \textbf{0.8420} & 0.8910 & \textbf{0.8635} & \textbf{0.9070} & 0.7637 & 0.8359  & \textbf{0.8480 } & \textbf{+0.0047} \\ \bottomrule
\end{tabular}

}
\label{tab:kuairand1k_performance}
\end{table*}

%% file: fig/adapter_comparsion.tex
%%%%%%%%%%%%%%%%%%%%%%%%%%%%%%%%%%%%%%%%%%%%%%%%%%%%%%%%%%
\begin{figure}[t!]
	\centering
	\includegraphics[width=\linewidth]{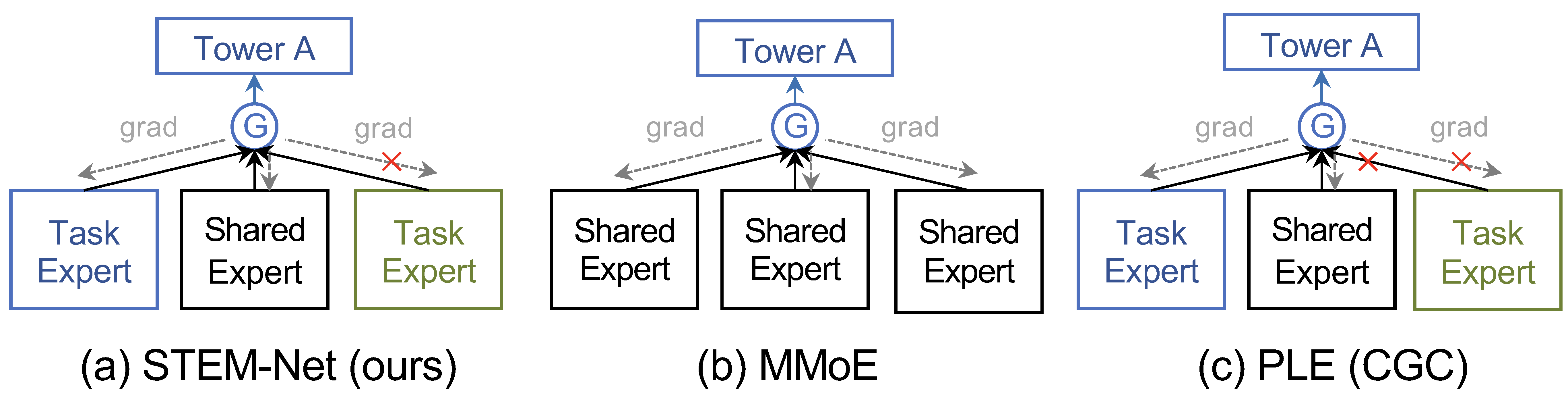}
	\caption{Comparison of gating networks of MMoE (All Forward All Backward), PLE (Task-specific Forward Task-specific Backward) and our STEM-Net (All Forward Task-specific Backward).}
	\label{fig:adapter_comparison}
\end{figure}
%%%%%%%%%%%%%%%%%%%%%%%%%%%%%%%%%%%%%%%%%%%%%%%%%%%%%%%%%%

%% file: sections/5_experiment.tex
% \section{Experiment}

\section{Performance Evaluation}\label{sec:performance}

\subsection{Experimental Setup}
\subsubsection{Public Datasets.} We choose three public datasets, namely TikTok, QK-Video~\cite{tenrec_2022}, and KuaiRand1K~\cite{tenrec_2022} for performance evaluation.
We replace features from all datasets that appeared less than 10 times in the training set by a default value.
The statistics of the processed dataset is presented in Table ~\ref{tab:datasets}.

\subsubsection{Baselines.} 
To establish a performance benchmark for comparison, we implement a \textbf{Single-Task} model, as well as popular MTL methods including \textbf{Shared-Bottom}~\cite{shared_bottom}, \textbf{OMoE}~\cite{mmoe}, \textbf{MMoE}~\cite{mmoe} and \textbf{PLE}~\cite{ple}. 
The Single-Task model adopts MLPs (Multi-Layer Perceptrons) that are trained only for a single task. In implementation, it is essentially equivalent to a Shared-Bottom model with only one tower. For the Tiktok and QK-Video datasets, we introduce two additional baselines: \textbf{ESMM}~\cite{ESMM} and \textbf{AITM}~\cite{AITMKDD2021}. 
Furthermore, we provide the Multi-Embedding versions of MMoE and PLE as strong baselines to study the effect of simply including additional embeddings, namely \textbf{ME-MMoE} (Fig.~\ref{fig:related_work}(d)) and \textbf{ME-PLE} (Fig.~\ref{fig:related_work}(e)). 
For ME-MMoE, we allocate independent embeddings for each expert, while for ME-PLE, we assign separate embeddings for task-specific and shared experts.
Note that ME-PLE also follows the STEM paradigm, in the sense that it also learns shared and task-specific embeddings. 
ME-MMoE and ME-PLE use the same gating network as MMoE and PLE, i.e., All Forward All Backward and Task-specific Forward Task-specific Backward, respectively.
\subsubsection{Hyper-Parameter Settings.} We implement all methods based on Pytorch and use Adam~\cite{adam} as the optimizer. 
We set the learning rate as $\{1e^{-3},5e^{-4},1e^{-4}\}$, the batch size as $4096$, and the $l_2$ regularization factor of embedding as $1e^{-6}$. We set the dimension of the embedding to 16, and each expert/bottom is an MLP with hidden units of $[512,512,512]$. 
The towers and the gate networks of all methods are MLPs with hidden units of $[128,64]$. The number of task-specific and shared experts is chosen from $\{1,2,4,8\}$. Grid search is used to find optimal hyper-parameters for all methods. 
\subsection{Overall Performance}
The comparison between STEM-Net and the baselines on the three datasets is presented in Tables~\ref{tab:tiktok_performance},~\ref{tab:qkvideo_performance} and ~\ref{tab:kuairand1k_performance}.We have the following observations.
First, STEM-Net achieves the best average AUC across all datasets.
Specifically, STEM-Net exhibits an average AUC improvement of approximately $4e^{-3}$, $2e^{-3}$, and $4e^{-3}$ over the state-of-the-art model (PLE) on three datasets, respectively. 

Second, STEM-Net achieves positive transfer over the single task model on the \texttt{Like} task on both TikTok and QK-Video dataset for the first time.
These tasks have much less positive feedback than the other tasks, making them fragile to negative transfer.
STEM-Net tackles negative transfer by learning task-specific embedding and hence capturing task-specific preference on this task.

Third, to make a fair comparison regarding number of parameters, we compare STEM-Net with another two multi-embedding baselines: ME-MMoE and ME-PLE.
STEM-Net and ME-PLE both have $T+1$ embedding tables, while ME-MMoE has the same number of embedding tables as experts (typically greater than $T$).
STEM-Net still beats these two methods on three datasets.
Furthermore, ME-PLE, which also follows STEM paradigm, outperforms ME-MMoE, especially on \texttt{Like} on TikTok and QK-Video datasets, validating the effectiveness of task-specific embeddings in capturing task-specific user preference.

\input{tables/sharedembedding_vs_taskspecific_embedding}

\subsection{Which feature fields should be task-specific?} 
We investigate which feature fields should be task-specific to learn diverse user preferences across tasks. 
To this end, we designed STEM-Net variants that only learn task-specific embeddings for selected fields $F$, denoted as STEM-Net-($F$).
Results on the TikTok dataset are presented in Table~\ref{tab:sharing_degrees}. 
% Experimental results on the TikTok dataset are presented in Table~\ref{tab:sharing_degrees} and Figure~\ref{fig:embedding_exp}. 

First, if no features are task-specific in STEM-Net, denotes as STEM-Net-$\varnothing$, its performance is comparable with that of MMoE and PLE. 
This proves that task-specific embeddings or the STEM paradigm (facilitated by All Forward Task-specific Backward gating) is the key for STEM-Net's performance lift, rather than the model architecture itself.

Second, our main purpose is to capture user's preference on items in STEM-Net, so we wonder if we can still achieve decent performance lift when only making embeddings of User ID and Item ID task-specific. 
We evaluate the performance of the corresponding model STEM-Net-(user id, item id), observing that it also achieves competitive performance with STEM-Net. 
This verify our hypothesis that STEM-Net's performance lift is mainly attributed to its ability to capture diverse user preference over items.

Further, we are curious whether user side (e.g., user id and device id in Tiktok) or item side (e.g., item id, author id, item city, channel, music id and video\_duration in Tiktok) features are more critical to have task-specific embeddings.
We design two new variants, STEM-Net-(user side) and STEM-Net-(item side), and observe the former one perform better, indicating that user side features are more effective than item in capturing diverse user preference among tasks.
\input{tables/adapter_abalation_study}

\input{fig/user_item_distance_analysis}

\subsection{Effectiveness of the Proposed Gating Network}
In STEM-Net, our proposed All Forward Task-specific Backward gating network is critical to learn task-specific embeddings via stop gradients (SG) operation.
Furthermore, the input to the gating network is also worth discussing~\cite{fei2021gemnn,chang2023pepnet}. 
Thus, we present three variants by replacing the input of Eq.~\ref{eq:gate}, i.e.,  $\bm{h}^{t}_{0}+\bm{h}^{S}_{0}$,  by $\bm{h}^{t}_{0}$ or $\bm{h}^{S}_{0}$, and conduct ablation study of SG on each variant. 
Experimental results are shown in Table~\ref{tab:gate_network}. Here are two observation:

(a) The stop gradient operation proves to be critical for performance lift. As shown in Table~\ref{tab:gate_network}, across all input settings, introducing SG (highlighted in gray) leads to an AUC improvement of approximately $5e^{-3}$ to $8e^{-3}$ for the \texttt{Like} task. Without SG, the embeddings are essentially shared between tasks, making the model similar to ME-MMoE.

(b) The best performance is achieved when the input consists of both task-specific and shared concatenated embeddings, resulting in an improvement of $1e^{-3}$ in average AUC. 
We argue this is because that task-specific embeddings are exclusively updated by their corresponding tasks, which restricts their ability to perceive the information of the other tasks. 
Consequently, integrating shared embeddings as a constituent of the input aids the gating network in perceiving common information across tasks.

\section{Contradictory User Preference Analysis}\label{subsec:contradictory}

We hypothesize that on the comparable subset where there is enough feedback from both tasks, there may be contradictory user preference over items across tasks. 
For example, a user $u$ may be inclined to an item $i$ regarding task \texttt{A}, but be declined to it or neutral regarding task \texttt{B}. 
For these user item pairs, shared embedding methods fail to learn such contradictory preference, since they have only a single shared embedding table, and therefore can learn \emph{only a single} preference.
In contrast, STEM-Net should be able to capture such contradictory preference by task-specific embeddings.

We conduct the follow analysis to validate the above hypothesis.
We select a set of contradictory user item pairs $S$ whose Euclidean distance\footnote{We choose Euclidean distance as the similarity metric because the experts adopt MLPs. } are among the top-40\% regarding single task \texttt{Like} embedding (Fig.~\ref{subfig:st_like_distance_distribution}) and among the bottom-40\% regarding single task \texttt{Finish} embedding (Fig.~\ref{subfig:st_finish_distance_distribution}).
These user item pairs correspond to 9.63\% of all samples.
We plot the distance distribution of these pairs regarding the shared embedding from PLE in Fig.~\ref{subfig:ple_distance_distribution} and observe that PLE learn small distances for most of them, which is similar to the distribution of single task \texttt{Finish}, while contradictory to that of single task \texttt{Like}.

In STEM-Net, the distance distribution of \texttt{Like} (Fig.~\ref{subfig:stem_like_distance_distribution}) and \texttt{Finish}-specific (Fig.~\ref{subfig:stem_finish_distance_distribution}) embedding table are consistent to single task correspondence, showing case that STEM-Net is able to learn the contradictory user item preference.
Note that similar to the PLE emebdding, the distance distribution of shared embedding table (Fig.~\ref{subfig:ple_distance_distribution}) in STEM-Net is also similar to the distribution of single task \texttt{Finish} while contradictory to that of single task \texttt{Like}.

\subsection{Online A/B Test}
\subsubsection{Online Deployment}
Since 2022, STEM-Net has been developed on Tencent's display advertising platform over various scenarios. These advertising scenarios consists of several tasks including \texttt{Follow}, \texttt{Activation}, \texttt{Fulfill Sheet}, and \texttt{Pay}. 

\begin{table}[t!]
\caption{AUC Lift of Online A/B Test}
\resizebox{\linewidth}{!}{
\begin{tabular}{l|ccccc}
\hline
Scenario          & Follow & Activation             & Fulfill Sheet             & Pay  & Avg.       \\ \hline
Scenario 1 & +0.29\%  & +0.33\% & +0.29\% & +0.35\% & +0.32\% \\
Scenario 2 & +0.13\% & +0.22\% & +0.33\% & +0.27\% & +0.24\%\\
Scenario 3 & +0.47\% & +0.39\% & +0.28\% & +0.78\% & +0.48\% \\
\hline
\end{tabular}
}
\label{tab:online}
\end{table}

\subsubsection{Performance}
The production model follows an MMoE architecture, where each expert is NFwFM, a variant of NFM~\cite{nfm2017} which replaces the vanilla FM~\cite{fm2010} by FwFM~\cite{fwfm}.
We equip the production model with shared and task-specific embeddings.
The overall improvements of all tasks are shown in Table~\ref{tab:online}, indicating a significant improvement of STEM-Net over the production MMoE model. 
In particular, STEM-Net brings 0.32\%, 0.24\%, and 0.48\% average AUC lifts for three representative scenarios, leading to 4.2\%, 3.9\%, and 7.1\% GMV lift in our online A/B test.

%% file: tables/sharedembedding_vs_taskspecific_embedding.tex
\begin{table}[]
\caption{Performance of STEM-Net variants where only selected feature fields deploy task-specific. 
}
\centering
\resizebox{\linewidth}{!}{
\begin{tabular}{l|cccc}
\toprule
\multicolumn{1}{c}{\multirow{2}{*}{Variants}} & \multicolumn{2}{c}{Tiktok} & \multicolumn{2}{c}{KuaiRand1K} \\
\multicolumn{1}{c}{} & AUC & \#Param & AUC & \#Param \\ \hline
STEM-Net-$\varnothing$ & 0.8261 & 1.00x & 0.8382 & 1.00x \\
STEM-Net-(user id, item id) & 0.8302  & 1.95x & 0.8448 & 1.36x \\
STEM-Net-(user side) & 0.8301 & 1.23x & 0.8437 & 1.00x \\
STEM-Net-(item side) & 0.8260 & 2.62x & 0.8193 & 2.05x \\
STEM-Net-(all features) & 0.8304  & 2.85x & 0.8480  & 2.06x \\ \bottomrule
\end{tabular}
}
\label{tab:sharing_degrees}
\end{table}

%% file: tables/adapter_abalation_study.tex
\begin{table}[tp!]
\centering
\caption{The effect of input and stop gradient to gating network.}
\resizebox{0.95\linewidth}{!}{
\begin{tabular}{@{}c|c|cc|cc|c@{}}
\toprule
\multirow{2}{*}{Input} & \multirow{2}{*}{SG} & \multicolumn{2}{c|}{Finish} & \multicolumn{2}{c|}{Like} & \multicolumn{1}{c}{\multirow{2}{*}{\begin{tabular}[c]{@{}c@{}}Average\\ AUC\end{tabular}}} \\
 &  & \multicolumn{1}{c}{AUC} & \multicolumn{1}{c|}{Logloss} & \multicolumn{1}{c}{AUC} & \multicolumn{1}{c|}{Logloss} & \multicolumn{1}{c}{} \\ \midrule
\multirow{3}{*}{$\bm{h}^{t}_{0}$} & \Checkmark & \cellcolor{gray!25}0.7509 & \cellcolor{gray!25}0.5107 & \cellcolor{gray!25}0.9077 & \cellcolor{gray!25}0.0555 & \cellcolor{gray!25}0.8293 \\
 & \XSolidBrush & 0.7511  & 0.5106 & 0.9023  & 0.0560  & 0.8267  \\
 & $\Delta$ & +0.0001 & -0.0001 & -0.0054 & +0.0005 & -0.0026 \\
 \hline
\multirow{3}{*}{$\bm{h}^{S}_{0}$} & \Checkmark & \cellcolor{gray!25}0.7510 & \cellcolor{gray!25}0.5106 & \cellcolor{gray!25}0.9089 & \cellcolor{gray!25}0.0553 & \cellcolor{gray!25}0.8300 \\
 & \XSolidBrush & 0.7511 & 0.5105 & 0.9037  & 0.0558  & 0.8274 \\
 
 & $\Delta$ & +0.0001 & 0.0000  & 0.0052 & +0.0005 & -0.0026  \\ \hline
\multirow{3}{*}{$\bm{h}^{t}_{0}+\bm{h}^{S}_{0}$} & \Checkmark & \cellcolor{gray!25}0.7513 & \cellcolor{gray!25}0.5104 & \cellcolor{gray!25}0.9095 & \cellcolor{gray!25}0.0553 & \cellcolor{gray!25}0.8304 \\
 & \XSolidBrush & 0.7510  & 0.5106 & 0.9008 & 0.0562 & 0.8259 \\ 
  & $\Delta$ & -0.0002 & +0.0003 & -0.0080 & +0.0009 & -0.0045 \\
 \bottomrule
\end{tabular}
}
\label{tab:gate_network}
\end{table}

%% file: fig/user_item_distance_analysis.tex
\begin{figure*}[t!]
	\centering  %图片全局居中
	\subfigbottomskip=0pt %两行子图之间的行间距
	\subfigcapskip=0pt %设置子图与子标题之间的距离
	\subfigure[Single Task (\texttt{Like} Embedding)]{
		\includegraphics[width=0.32\linewidth]{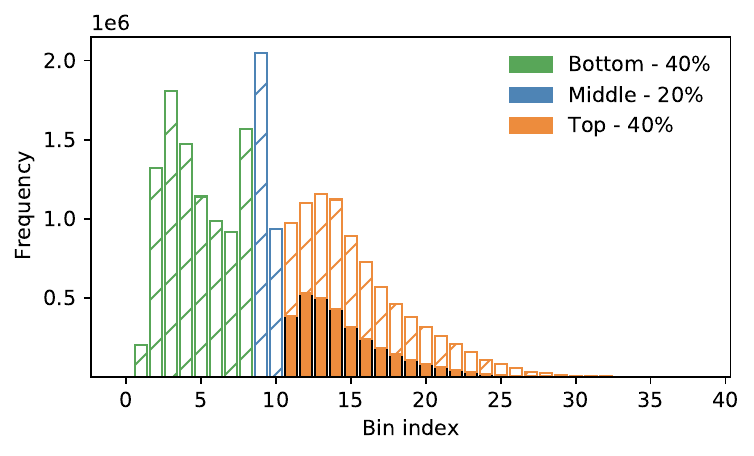}
            \label{subfig:st_like_distance_distribution}}
	\subfigure[Single Task (\texttt{Finish} Embedding)]{
		\includegraphics[width=0.32\linewidth]{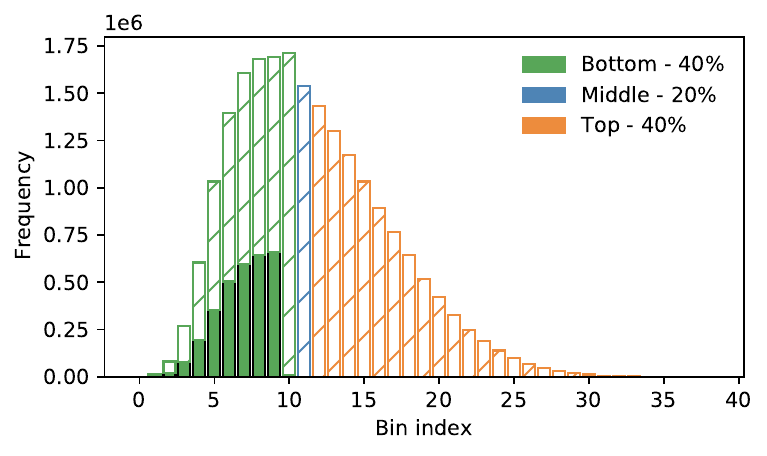}
            \label{subfig:st_finish_distance_distribution}}
	\subfigure[PLE (Shared-Embedding)]{
		\includegraphics[width=0.32\linewidth]{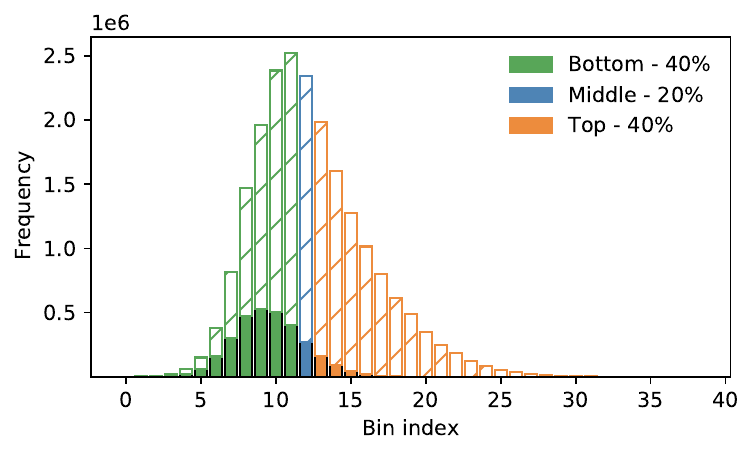}
            \label{subfig:ple_distance_distribution}}
	\subfigure[STEM-Net (\texttt{Like} Embedding)]{
		\includegraphics[width=0.32\linewidth]{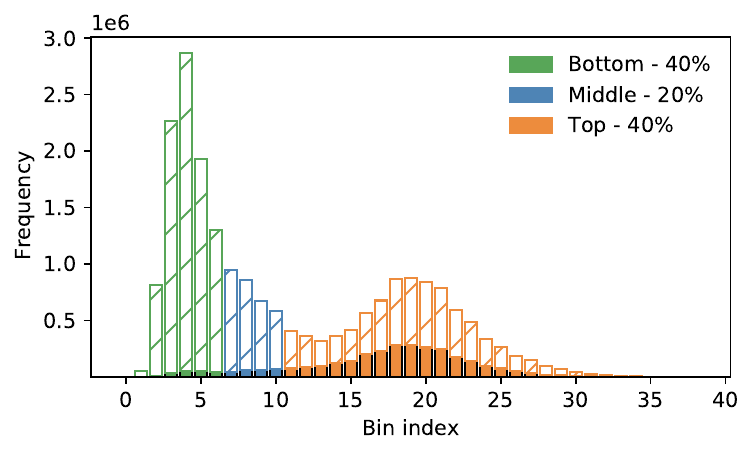}
            \label{subfig:stem_like_distance_distribution}}
	\subfigure[STEM-Net (\texttt{Finish} Embedding)]{
		\includegraphics[width=0.32\linewidth]{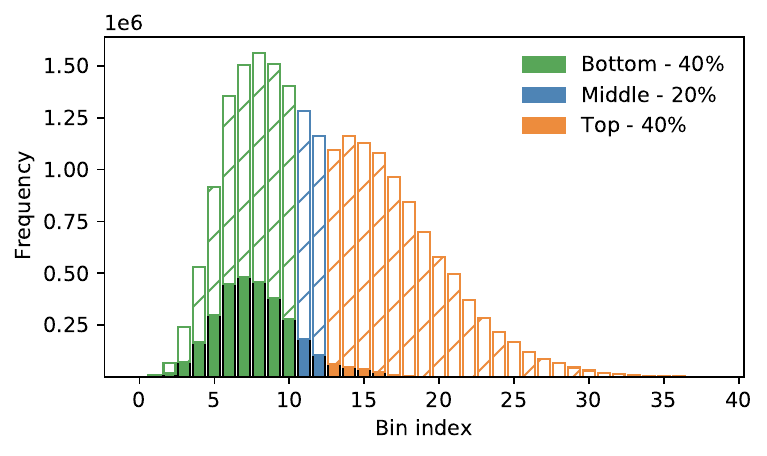}
            \label{subfig:stem_finish_distance_distribution}}
	\subfigure[STEM-Net (Shared Embedding)]{
		\includegraphics[width=0.32\linewidth]{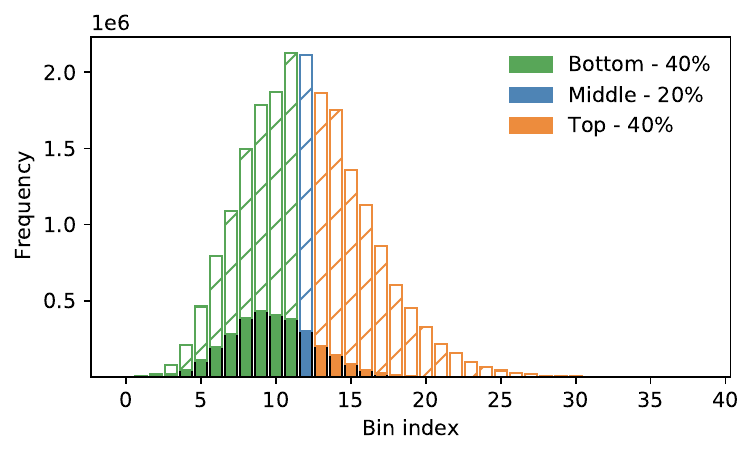}
            \label{subfig:stem_shared_distance_distribution}}
	\caption{The distance distribution of the contradictory user item pair set $S$ (with solid color) as well as the whole user item pair set (with slash lines) regarding: the single task \texttt{Like} (a) and \texttt{Finish}  embedding (b), the PLE embedding (c), and the \texttt{Like} (d) and \texttt{Finish}-specific (e) embedding and shared embedding  (f) in STEM-Net.}
 %The distribution distances (\textit{i.e.}, the \textit{l2} norm of the difference) between user embeddings and item embeddings in the task conflict subset of Tiktok dataset, which is constructed based on the intersection of the samples that have similar embeddings in \textit{finish} single-task, while distinct embeddings in \textit{like} single-task. We use the 40th and 60th percentiles of the whole samples as thresholds to discriminate "Similar", "Moderate" and "Distinct". \# means "the number of". The colored bars represent the samples in the task conflict subset, while the diagonal lines represent samples not in this subset. The first row presents the distance distribution for single-task models for the like and finish tasks, as well as the shared-embedding multi-task model (PLE). The second row presents the distance distribution for task-specific embeddings and shared embeddings in STEM.}
	\label{fig: embedding_distance_analysis_finish_similar}
\end{figure*}

%% file: sections/6_conclusion.tex
\section{Conclusion}
In this paper, we propose a novel Share and Task-specific EMbedding paradigm to tackle the negative transfer in MTL recommendation.
We design a simple model under such paradigm, namely STEM-Net, which demonstrates compelling performance on comparable samples, achieving positive transfer.
In three public datasets and industrial online A/B test, we validate that our proposed STEM-Net achieves significant performance lift over state-of-the-art MTL recommendation models.